\documentclass[twocolumn,showpacs,aps,prl,a4paper]{revtex4}
\usepackage{amsmath,graphicx}

\begin{document}

\title{Atomic and molecular matter fields in periodic potentials.
}
\author{Klaus M\o lmer}
\affiliation{
QUANTOP, Danish National Research Foundation Center for Quantum
Optics, Department of Physics and Astronomy,\\
University of Aarhus \\
DK-8000 \AA rhus C, Denmark}

\begin{abstract}
This paper deals with the conversion between atoms and molecules
in optical lattices.  We show that in the absence of
collisional interaction, the atomic and molecular components in different 
lattice wells combine into states with macroscopic condensate
fractions, which can be observed as a strong  
diffraction signal, if the particles are abruptly released 
from the lattice.  The condensate population, and the  diffraction signal
are governed not only by the mean number
of atoms or molecules in each well, but by the precise  amplitudes on
state vector components with different
numbers of particles. We discuss ways to control
these amplitudes and to maximize the condensate
fraction in the molecular formation process.
\end{abstract}


\maketitle

\section{Introduction}

The immersion  of degenerate gases in laser induced periodic
potentials has led to numerous beautiful studies of 
both many-body and few-body phenomena, including squeezing, 
the superfluid-insulator phase transition, and
collisionally induced collapses and revivals of 
the phase of a quantum field with few quanta (atoms)
\cite{Yalediff,Munichdiff,Munichrevival}.
We have previously shown \cite{Ess03} that with an
average population of approximately a single atom per site or less, 
the  removal of atom pairs from the lattice may be used to
reestablish a robust atomic interference pattern even long times 
after dephasing has taken place due to collisional interaction among the
atoms. This works, because the remaining zero and one-atom components of the 
quantum state have experienced no interaction, and hence
no dephasing.  In the same work we showed that if this removal occurs by 
formation of a bound molecular state, a molecular condensate 
may be produced and detected when the lattice is removed. 
The incommensurate Rabi frequencies
governing the dynamics among different population components
present a major obstacle towards extending the results to the case 
with more particles in every lattice well, and
the restriction to  superposition states dominated by
a vacuum component and a single atom and a single molecule component 
on every lattice site was crucial for the analysis in \cite{Ess03}.

In the present paper we first motivate our work by a brief 
discussion of the recent experimental and theoretical work 
on formation of molecules in cold atomic gases.
We then summarize the analysis
of the connection between the quantum states of particles in indivual wells
of a periodic potential, 
diffraction patterns, and the macroscopic population of a single condensate
wave function. Finally,
we turn to the process of converting more than a single pair of atoms
into molecules in every lattice site, and we present a method to do
this with maximal possible coherence of the final state.
We shall not make use of atomic interactions
and of the Mott-insulator phase
transition dynamics, which can prepare an exact integer number of atoms
per site and hence also 
an exact number of molecules \cite{Jaksch02,Damski}.
We assume instead that the lattice potential is applied to
the condensate too fast for the 
Mott insulator transition to happen, but slowly enough that the
spatial motion of the individual atoms and molecules  follow the 
lowest energy state of the potential.

\section{Conversion between trapped atoms and molecules}

By carefully designed photoassociation
\cite{Heinzen00}, which makes use of the
fact that the two-atom component is in a fully determined initial
state, it is possible to convert atoms in a condensate effextively into 
molecules. By tuning a magnetic field across an atomic Feshbach scattering 
resonance in $^{85}Rb$, Donley et al \cite{Donley} observed a significant
production of energetic atoms, and a loss of atoms, which could
be well accounted for by the production of a loosely bound molecular
system \cite{Holland, Burnettmol, Greene}. Also thermal
\cite{chu03} and condensed \cite{grimm03} $^{133}Cs$ atoms have been
converted into molecules by tuning of a magnetic field, and 
the fermionic atoms $^{40}K$ and $^6Li$  have been converted into
bosonic diatomic molecules \cite{regal03,Salomon03}.
The long lifetimes of molecular states formed by
fermionic atoms \cite{Petrov03}
suggest unique possibilities to study condensation
of the molecular component and to convert the system back to the atomic
system in the BCS superfluid state \cite{Castin03}.

All of these above mentioned experiments and theories deal with 
large quasi-homogeneous systems, for which mean field theories 
are well justified, and for which the atoms and molecules, ideally,
populate  only a single spatial  quantum state. It has been proposed
to study the molecular formation processes after the atoms have been 
subject to a periodic potential in form of an optical lattice
\cite{Jaksch02,Damski,Ess03,Molmer03}. 
The clustering of the atoms in the potential
wells increases significantly the two-atom density correlations
of the system and hence the molecular formation rate. In addition, the
quantization of the motion in the narrow wells, makes
the photoassociation process a transition between well-reseolved 
discrete states
with resulting better possibilities to control the coherence properties,
and avoid 'rogue' photoassociation \cite{Javanainen}.
From an experimental point of view our molecules
can be prepared in a deeper bound molecular state,
selected coherently by the coupling laser fields.
In \cite{Jaksch02,Damski} it is proposed to invoke the Mott-insulator
dynamics and prepare a system with precisely two atoms per well,
to turn these atoms into a single molecule, and to subsquently melt the
molecular Mott insulator and form a molecular condensate. In 
\cite{Ess03,Molmer03}, we do not apply the Mott dynamics, but
we rather accept the fluctuating atom number in each well, and use the
fluctuations
as a seed to prepare  mean molecular fields in every
single well, which combine to a macroscopic condensate upon
release of the atoms. Cold atoms in lattices offer a number of 
interesting possibilities for many-body physics, quantum optics 
and quantum information studies, and we believe that the conversion
to molecules will provide extra possibilities to study multi-component
dynamics, and, e.g., mixed Bose-Fermi systems, and for detection of 
low-temperature phase transition dynamics.

\section{Diffraction and condensate fraction in a lattice system}

Apart from the restriction of the atomic spatial dynamics induced by the
confining potential, the periodicity is the most characteristic feature
of the optical lattice, and in this section we shall summarize
some simple properties related to this periodicity.

\subsection{Splitting of number states and coherent states}

We assume that we have initially a quasi-homogeneous
Bose-Einstein condensate at rest. This condensate 
may contain a definite number $N$ of atoms, which
fragments with a probabilitity distribution $p(n)$ for the 
occupancy  $n$ of each of a large number of wells. The state of the
system is described by a quantum mechanical many-body wave function,
and the contents of each well is necessarily entangled
with the contents of the other wells. 
If, instead the entire many body wavefunction is described as a coherent
state, i.e., an eigenstate of the atomic annihilation operators
$\hat{\Psi}(x)$, in the absense of interaction among the atoms, this 
state factors
{\it exactly} into a product state of coherent states populating each well.
Each site is in that case populated by a superposition 
$\sum_n c_n|n\rangle$ of zero, one, or
more atoms, where  the amplitudes $c_n$ can be
parametrized by a single coefficient $\beta$:
$c_n=\exp(-|\beta|^2/2)\beta^n/\sqrt{n!}$. As long as the lattice
potential is kept deep enough, the factorization of the 
total system into the contents of each well is maintained in the future
evolution of the system, and hence we need only determine the state
vector evolution for the combined atom-molecule system in each well,
and finally obtain the macrocopic properties from the resulting product
state.

The macroscopic coherent
state is a quantum mechanical superposition of states with a total
atom number $N$ distributed according to a Poisson distribution, and
with a specific relative phase $\phi$ between different number state
components.  Such a superposition state does not have a definite total
number of atoms, but the number fluctuations are 
insignificant in comparison with the total number $N$ in the 
limit of large condensates, and 
one will not see any difference
between predictions based on the coherent state and on a 
state with precisely $N$ atoms. Remarkably, this argument also
applies when the condensate is being split over many wells,
where the relative fluctuations are {\it not} insignificant.
The reason is that the poissonian fluctuations in atom number on each lattice 
site account properly for the entanglement of the content of the site with
the remanining system, and the relative phase between number states
assumed by the coherent state Ansatz leads to correct predictions for 
the interference between different wells, even though
this interference is due to entanglement
rather than mean-fields \cite{Theorydiff,Klausdiff}.  A more formal argument
states that a Poissonian mixture of number states is identical to
a mixture of coherent states with random phase and the same mean
population, and hence predictions for the outcome of any time evolution 
and subsquent measurement on states picked at random from either
ensemble will be strictly identical \cite{Klausdiff}, also when this
evolution involves splitting the system in smaller components. We shall 
apply the coherent state Ansatz throughout the paper.

\subsection{Atomic and molecular field operators in position,
momentum and Wannier state representation}

When the optical lattice is applied slowly to the zero momentum
condensate, it evolves into the the zero momentum state
of the lowest Bloch band of the potential, which is, in case of a
deep potential modulation, a real superposition of Wannier states
localized in each well, $\phi_m(x)=\phi_0(x-mL)$.

If the atoms occupy only the ground states of the wells of the lattice,
it is convenient to introduce the discrete set of atomic field
operators, $a_m$,  that remove atoms from the Wannier mode functions 
rather than $\hat{\Psi}(x)$, which remove atoms from 
specific locations in space,
\begin{equation}
a_m = \int \phi_m(x) \hat{\Psi}(x) dx.
\end{equation}
The mode operators obey the standard commutator relations
$[a_{m},a^{\dagger}_{m'}]= \delta_{m,m'}$.

Let $|n\rangle_m$ denote the state with $n$ atoms populating the 
spatial wavefunction $\phi_m(x)$.
In the state
$\sum_n c_n|n\rangle_m$, the mean value of 
the mode annihilation
operator in the well equals
$\alpha := \langle a_m\rangle = \sum_{n=1}^{\infty}
c_{n-1}^*c_n\sqrt{n}$, the mean
number of atoms equals 
$\overline{n} := \langle a^{\dagger}_ma_m\rangle=
\sum_{n=1}^{\infty} |c_n|^2 n$,
and in the product state  $\Psi\rangle = \Pi_m(\sum_n  c_n|n\rangle_m)$
we have the inter-well coherence
$\langle a_{m'}^{\dagger}a_m\rangle = \alpha^*\alpha$ for $m \ne m'$.

The condensate population is defined as the largest eigenvalue
of the one-body density matrix
$\langle\hat{\Psi}^{\dagger}(x)\hat{\Psi}(x')\rangle$, and in the basis of 
localized Wannier
states, the mean occupancy  $\overline{n}$ of the wells appears 
in the diagonal and
the squared mean field amplitude $|\alpha|^2$ appears in all other 
positions in
this matrix. In a system with $N_L$ wells,
the eigenvector with equal amplitude on each Wannier function has the
largest eigenvalue of $(N_L-1)|\alpha|^2+\overline{n}$, i.e., in the limit of
large $N_L$ the condensate fraction is $|\alpha|^2/n$, in agreement
with the value predicted by the
off-diagonal long-range order, $\langle a^{\dagger}_ma_{m'}\rangle
= |\alpha|^2 = \overline{n}\cdot(|\alpha|^2/\overline{n})$. 
This analysis of condensate population
and fraction also applies to the molecules if $\overline{n}$ and $\alpha$ are
replaced by the molecular mean population $\overline{n}_M$ and 
the molecular mean field $\alpha_M$.
Note that if the potential is removed slowly,
the many fragments will combine together to form a single condensate
with the population just identified.

In \cite{Ess03}, we introduce  $\hat{\Psi}(k) =
\frac{1}{\sqrt{2\pi}}\int dx e^{ikx}
\hat{\Psi}(x)$,
and we present a derivation
of the momentum distribution of the particles after release from the periodic
potential, 
\begin{eqnarray}
\label{pattern}
\langle \hat{\Psi}^{\dagger}(k) \hat{\Psi}(k)\rangle \nonumber \\
 = N_L|\phi(k)|^2\left( (\overline{n}-|\alpha|^2)+
|\alpha|^2\sum_q \delta(k-q\frac{2\pi}{L})\right),
\label{distribution}
\end{eqnarray}
where $\phi(k)$ is the Fourier transfom of the Wannier function $\phi_0(x)$.
The sum over integers $q$
gives rise to a comb at lattice momenta $q\frac{2\pi}{L}$
with a modulation proportional to the square of the mean field
amplitude $|\alpha|^2$. This comb sits on top of a flat background caused by
the incoherent population $\overline{n}-|\alpha|^2$ of the wells, 
and the whole distribution
is comprised within the width of the single well momentum distribution
$|\phi(k)|^2$.

Using a lattice for the 
study of this system is not only convenient for the tayloring of 
the association process, or for the possibility to observe Mott
insulator dynamics: the multi-well fragmentation of the system
leads to a diffraction pattern with a clear identification of the coherence 
properties and the condensate fraction of the system. 
By counting the released
atoms or the molecules on a position sensitive detector the values
of $\overline{n}$ and $\alpha$ and thus the condensate fraction of both atoms and
molecules can be determined.

\section{Converting several atoms into molecules}

We now turn our attention to the formation of molecules.
The above section provides the necessary connection
to the macroscopic signals and condensate properties, and we shall hence
treat the dynamics in a single well, only.

\subsection{Theoretical maximum values}

Let us first assess the theoretical maximum number of molecules
and maximum mean molecular field, in the case where an atomic
state $\sum_n c_n|n\rangle$ is made subject to the association
process. For even $n$ it is possible to remove the entire atomic
component and create $k=n/2$ molecules, whereas for
odd $n$ one may obtain $k=(n-1)/2$ molecules, and one is left
with a single atom. We can write the maximum number of molecules
for each atomic number state component as $k=(n/2-1/4)+(-1)^n/4$, and thus
obtain the maximum expectation value of the number of molecules,
\begin{equation}
\langle n_M\rangle \leq \frac{\langle n\rangle}{2}-\frac{1}{4}+
\frac{1}{4}\langle (-1)^n\rangle.
\end{equation}
For a coherent state with mean atomic population $\overline{n}$,
we get 
\begin{equation}
\langle n_M\rangle \leq  \frac{\overline{n}}{2}-\frac{1}{4}+
\frac{1}{4}e^{-2\overline{n}},
\end{equation}
and for a thermal state (exponential
distribution) with mean atomic population $\overline{n}$,
we get 
\begin{equation}
\langle n_M\rangle \leq  \frac{\overline{n}}{2}-
\frac{1}{4}\frac{2\overline{n}}{2\overline{n}+1}.
\end{equation}

The mean molecular
field amplitude is maximized if the atoms are maximally
converted to the state
\begin{equation}
|\Psi\rangle = \sum_k r_{2k} |k\ mol.\rangle + 
\sum_k r_{2k+1} |1\ atom,k\ mol.\rangle,
\end{equation}
and if the amplitudes on the molecular states 
are all real and positive ($r_n=|c_n|$), so that we obtain 
$\alpha_M=\sum_k (r_{2k-2}r_{2k}+r_{2k-1}r_{2k+1})\sqrt{k}$.
We have not obtained a closed analytical form for this maximum, but
for a coherent atomic input state we find 
that the square of the maximal mean molecular
field approaches $|\alpha_M|^2 \sim \langle n_M\rangle - 1/8$
in the limit of large $\overline{n}$.

\subsection{Numerical analysis}

\begin{figure}[h]
  \centering
 \includegraphics[width=6cm,angle=270]{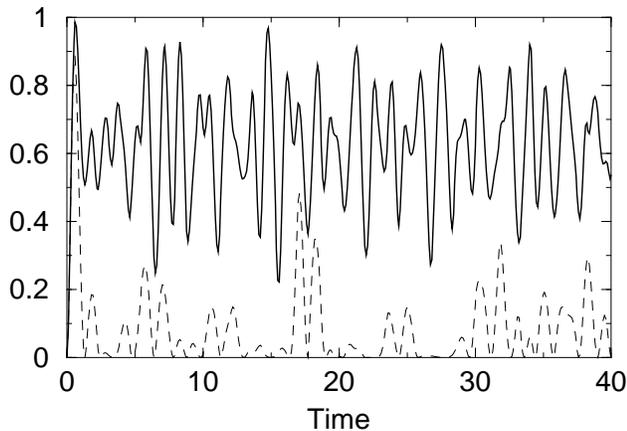}
\caption{Incoherent and coherent molecular components prepared by
photoassociation of atoms. The number of molecules (solid line)
and the squared norm of the mean molecular field (dashed line) as
a function of time are shown, starting at time zero with a coherent
atomic state with a mean number of atoms of 3. Time is measured in
units of the reciprocal coupling constant $\chi^{-1}$.}
\label{figure}
\end{figure}

The above quoted results are theoretical maximum values. The purpose of this
section is to investigate how close it is possible to arrive to these
values in an experiment.
We have solved numerically the
time evolution of the simple photoassociation Hamiltonian
\begin{equation}
H_{PA} = \chi(a^2b^{\dagger}+(a^{\dagger})^2b)
\label{HPA}
\end{equation}
where $a(b)$ is the operator of annihilation of an atom (molecule) at an
arbitrary site. The site index is suppressed for convenience, since
exactly the same process takes place at every site in the lattice. 
Starting with a coherent atomic state, the Hamiltonian
(\ref{HPA}) introduces in every lattice well a superposition state
$\sum c_{n,n_M}(t)|n,n_M\rangle$, from which the mean number of
molecules and the mean molecular field is readily calculated. 
Atom-atom interactions and interactions between the
atoms and molecules are readily incorporated in these calculations,
but they are set to zero in the present study, since our main focus is
on processes which are fast on the time scale of interactions.

Fig. 1  shows the results of a calculation with a constant resonant coupling.
The initial atomic state is a coherent state with $\overline{n}=3$ atoms
per well.
The figure shows that, at first, atoms are effectively converted into
molecules, but the conversion stops before the maximum possible value
of 1.251 molecules per site is achieved, and the square of the mean field does
not reach its maximum which we have computed to be 1.165. 
The long time  dynamics shows a pattern
of collapses and revivals of large amplitude oscillations, linked with 
the different
Rabi frequencies between the different number components of the quantum
state of the system.

\subsection{Adiabatic passage}

\begin{figure}[h]
  \centering
 \includegraphics[width=6cm,angle=270]{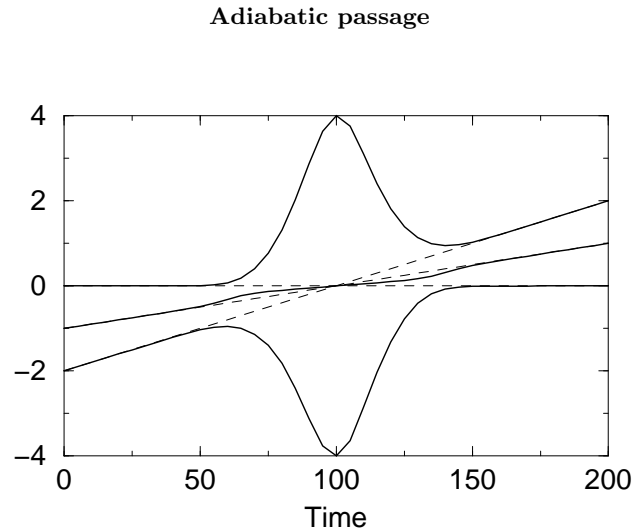}
\caption{Adiabatic eigenstate energies during chirped photoassociation
of the atoms.  The dashed lines in the figure
show the bare state energies of states with  4 atoms, 2 atoms and one molecule,
and 2 molecules in units of the maximum Rabi frequency $\chi$ during the
chirp. The line with the largest slope correponds to the purely
atomic component which is initially below and finally above resonance
with the association process. The solid curves are the eigenvalues of the
Hamiltonian, and we observe that the purely atomic state is continuously
transformed into the purely molecular state, which is the lowest energy 
state after the chirp. Similar spectra exist for all values of the initial
atomic population. Due to the factors $\sqrt{n}\sqrt{n-1}\sqrt{n_M+1}$
in the coupling of $n$ atoms and  $n_M$ molecules  to the state with
$n-2$ atoms and $n_M+1$ molecules, the splitting at resonance differs
for the different number state components.}
\label{figure}
\end{figure}

In order to increase the number of molecules, we propose to 
apply a frequency chirp to the photo association lasers
across the resonance. If this chirp is slow enough, the 
system will follow adiabatically the eigenstates of the 
time dependent Hamiltonian. If the lasers are initially detuned
to a too low energy, the photon dressed state with $n$ atoms has lower energy
than the state with $n-2$ atoms and one molecule etc., whereas at the
end of the chirp, these are the states with the highest energies, and
the lowest energy state is the one with none or a single atom, and $n/2$
or $(n-1)/2$ molecules. The chirp applies to all atomic $n$-components
at once, and if the chirp is slow, the system follows the
lowest energy eigenstate throughout the process, and the atoms  are all 
converted into states with the largest possible number of molecules.
Fig. 2. shows the energy eigenstates during a chirp, where the coupling
strength $\chi(t)$ is turned on and off as a sech-function of time, 
and the detuning
initially takes a value which is minus the maximum coupling Rabi frequency
$\chi_{max}$, and which is 
scanned linearly in time to a final value equal to  the maximum
coupling Rabi frequency. The plot shows the states with 4 atoms, 2 atoms and
a molecule and 2 molecules, which are coupled by the lasers. Similar spectra 
exist for all other initial number states for the atoms.

\begin{figure}[h]
  \centering
 \includegraphics[width=6cm,angle=270]{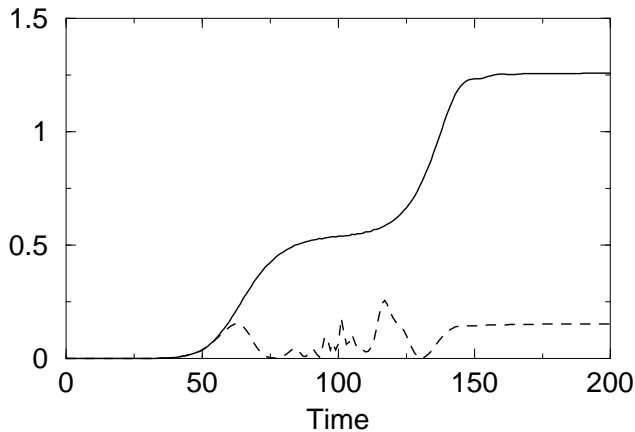}
\caption{Number of molecules (solid line) and square of mean molecular field
(dashed line) as functions of time during adiabatic chirp across the
photo association resonance. The initial atomic state is a coherent state
with a mean population of three atoms per well. The molecular population
at the end of the chirp is within reach of the theoretical maximum, but the
mean field is clearly much smaller than we could have hoped for.}
\label{figure}
\end{figure}

An example of the results of this process
is shown in Fig. 3, where we observe that more molecules are produced
than in the case of constant couplings shown in Fig. 1. Time
is given in units of the maximum value of the time-dependent coupling
coefficient $\chi_{max}^{-1}$.

The chirp may be carried out in different ways, and we do not claim to
have identified the optimum. 
The figure shows that we produce close to the maximally expected number
of molecules but it also shows, that the coherent component is very
weak.  As discussed in \cite{Munichdiff}, the short range interaction
between atoms causes a phase evolution
of the $c_n$ amplitudes with a frequency $\frac{U}{2}n(n-1)$,
leading to a reduction
of the atomic field amplitude $\alpha$  since the terms
$c_{n-1}^*c_n\sqrt{n}$ acquire different complex phases.
The resulting disappearance of atomic interference has been observed
in experiments \cite{Munichdiff,Yalediff}, and subsequent revivals
when the phase differences reach multiples of $2\pi$  have also been 
shown \cite{Munichrevival}. 
In our case, the loss of mean field is 
a consequence of the different complex phases 
attained by  the the different state
vector amplitudes as a result of the association process. The adiabatic
process, following the lower solid curve in Fig. 2., implies that a phase
factor proportional to the area between this curve and the horizontal line
will appear on the molecular state. This area, end hence the phase, will
be different for the adiabatic states obtained with other number state
components.
If the lattice is removed at the end of the process 
depicted in Fig. 3., the molecules will populate a number of different 
spatial states, 
and the zero momentum macroscopic state will only account for about 1/8 of
the molecules. The phases are of course determined by our computation,
but it is a non-trivial task to act on the system 
and individually control these values.  
We shall instead present a strategy, inspired by
composite pulse and spin echo techniques applied in nuclear magnetic resonance
studies \cite{Levitt},
that accomplishes full phase control of the system. 

\subsection{Cold atoms $\rightarrow$ 'hot' molecules  $\rightarrow$
'hot' atoms $\rightarrow$ 'cold' molecules}

Looking at Fig. 2., we note that
the unknown phases are simply given by the areas between the horizontal 
line and the time dependent energy levels. Observing that the 
time dependent energy spectra
in the figure are symmetric with respect to inversion in the crossing of the
dahsed diabatic lines, we can thus
identify a means to cancel these phases: After one chirp, the purely
atomic components have followed the states with the 
lowest energy to become components with zero or one atom and
a maximum number of molecules. If one repeats the chirp, on the system,
which now starts with molecules, i.e., the upper state on the left,
the system will now follow the {\it highest} energy state and return
to the purely atomic component, which is the highest state
in the right part of the figure. The accumulated phase in the second
chirp will exactly annihilate the one of the first chirp, and since this
is true for all initial atomic number states, the combination
of the  two chirped intractions  do
not change the state of the system. Now, let us instead perform the
second chirp at half the rate of the first one. In this case, the
phases accumulated on all components are twice the negative of 
the phases of the 
first chirp. The atoms have been recreated, but they have picked up
phases so that the atomic system is actually not coherent any more, and if
it is released at this point, its condensate fraction will be poor. 
We are not finished
yet, however, and performing a third chirp on the system at the
original chirp rate, we again turn as many atoms as possible into 
molecules, and the phases of the two rapid chirps and the single slow chirp
will cancel to produce a real linear combination of molecular
states.

\begin{figure}[h]
  \centering
 \includegraphics[width=6cm,angle=270]{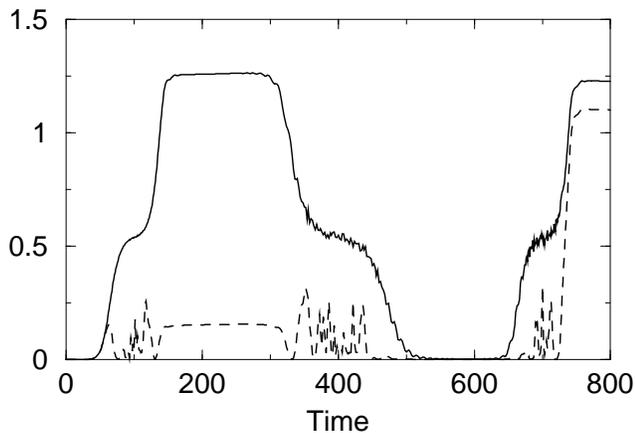}
\caption{Number of molecules (solid line) and square of mean molecular field
(dashed line) as functions of time during three adiabatic chirps across the
photo association resonance. The initial atomic state is a coherent state
with a mean population of three atoms per well. The first chirp, of duration
$t=200\chi_{max}^{-1}$ is as depicted in Fig. 3. The second chirp with
duration $t=400\chi_{max}^{-1}$ converts the molecules back into atoms,
and the last chirp produces the molecules again, but now with a sizable
mean field, as shown by the dashed line.  }
\label{figure}
\end{figure}

We have simulated this process numerically, and 
Fig. 4. shows the results of three-chirp photo association
pulses applied to the atomic system with an initial
mean population of three atoms per well. The figure shows that after
the first chirp, we have many molecules, but a small molecular
mean field. After the second chirp we have no molecules left,
and after the third chirp, the molecules are restored and now
with appropriate phases to ensure a strong molecular mean field.

\begin{figure}[h]
  \centering
 \includegraphics[width=6cm,angle=270]{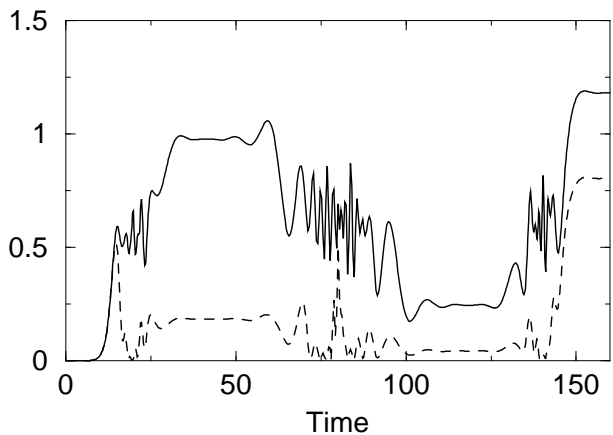}
\caption{Same as Figure 4., but with 5 times faster chirps.
The outcome is not as good as in Figure 4, but still better than
the results of a constant coupling, shown in Fig. 2.}
\label{figure}
\end{figure}

Fig. 5. shows the results of the same sequence but with faster chirps.
Both the total and the coherent molecular population are reduced compared
to the slow chirps, and a more detailed calculation is necessary to
find optimum chirp schemes, e.g., as function of the total duration
available for an experiment.

\section{Discussion}

In summary, we have presented a method 
to prepare a molecular system
in a macroscopicically populated quantum state.
The molecular condensate
exists due to the simultaneous existence on every lattice site
of different number state components, which are inherited
from the number fluctuations in the atomic system. The system
is too small to be treated by a mean field Ansatz,
and a separate derivation of the mean value of the field
operator indeed shows that the phase is ill-defined. Due to the
deterministic character of the problem - there is no decoherence
or loss at play - we were able to apply a controlled interaction,
in which molecules were first created, then removed, and finally
created again with a well-behaved mean field. This technique, and
possible rapid versions which utilize
composite pulse sequences, may be interesting to study in their own
right for this system, and we believe that they may also be 
highly relevant in other studies with optical lattices with
small occupation numbers on each site, e.g., in connection with 
simulations of quantum magnetism and superfluidity.

\noindent
{\it Note added --.} After submission of this paper, the first 
experimental results were reported on observation of molecular
condensates, formed by association of fermionic potassium and
lithium atoms \cite{latest}.


\begin{thebibliography}{99}

\bibitem{Yalediff} C. Orzel, A. K. Tuchman, M. L. Fenselau, M. Yasuda,
and M. A. Kasevich, Science {\bf 291}, 2386 (2001).

\bibitem{Munichdiff} M. Greiner, O. Mandel, T. Esslinger, T.W. H\"ansch, and
I. Bloch, Nature, {\bf 415}, 39-44 (2002).

\bibitem{Munichrevival} M. Greiner, O. Mandel, T. W. H\"ansch, and
I. Bloch, Nature, {\bf 419}, 51-54 (2002).

\bibitem{Ess03}
T. Esslinger and K. M\o lmer,
Phys. Rev. Lett. {\bf 90}, 160406 (2003).

\bibitem{Jaksch02} D. Jaksch, 
V. Venturi, J. I. Cirac, C. J. Williams, and P. Zoller,
Phys, Rev. Lett. {\bf 89}, 040402 (2002).

\bibitem{Damski} B. Damski, L. Santos, E. Tiemann, M. Lewenstein, S.
Kotochigova, P. Julienne, and P. Zoller, Phys. Rev. Lett. {\bf 90}
, 110401 (2003) 

\bibitem{Heinzen00} R. Wynar, R.S. Freeland, D.J. Han, C. Ryu, and 
D.J. Heinzen, Science {\bf 287}, 1016 (2000).
C. McKenzie, J. Hecker Denschlag, H. H\"affner, A.
Browaeys, Lus E. E. de Araujo, F. K. Fatemi, K. M. Jones, J. E.
Simsarian, D. Cho, A.  Simoni, E. Tiesinga, P. S. Julienne, 
K. Helmerson, P. D. Lett, S. L. Rolston, and W. D. Phillips,  
Phys. Rev. Lett. {\bf 88}, 120403 (2002).

\bibitem{Donley}
E. A. Donley, N. R. Claussen, S. T. Thompson, and C. E. Wieman,
 Nature {\bf 417}, 529 - 533 (2002).

\bibitem{Holland}
S.J.J.M.F. Kokkelmans and M.J. Holland, Phys. Rev. Lett. 
{\bf 89}, 180401 (2002).

\bibitem{Burnettmol} T. Koehler, T. Gasenzer, and  K. Burnett,
Phys. Rev. A {\bf 67}, 013601 (2003).

\bibitem{Greene} B. Borca, D. Blume, and C. H. Greene, cond-mat/0304341.

\bibitem{chu03} C. Chin, A. J. Kerman, V. Vuletic, and S. Chu, 
Phys. Rev. Lett. {\bf 90}, 033201.

\bibitem{grimm03} J. Herbih, T. Kraemer, M. Mark, T. Weber, C. Chin, H.-C.
N\"agerl, and R. Grimm, ScienceExpress, published online 21 August 2003;
10.1126/science.1088876/DCI.

\bibitem{regal03} C. A. Regal, C. Ticknor, J. L. Bohn, and D. S. Jin,
Nature {\bf 424} 47 (2003).

\bibitem{Salomon03} J. Cubizolles, T. Bourdel, S. J. J. M. F. Kokkelmans,
G. V. Shlyapnikov, and C. Salomon, cond-mat/0308018.

\bibitem{Petrov03} D. S. Petrov, C. Salomon, and G. V. Shlyapnikov,
cond mat/0309010.

\bibitem{Castin03}  L.D. Carr, G.V. Shlyapnikov, and Y. Castin, cond-mat/0308306.

\bibitem{Molmer03} K. M\o lmer, Phys. Rev. Lett. {\bf 90}, 110403 (2003).

\bibitem{Javanainen} M. Mackie, Ryan Kowalski, and Juha Javanainen,
 Phys. Rev. Lett. {\bf 84}, 3803
(2000); J. Javanainen  and M. Mackie, Phys. Rev. Lett. {\bf 88}, 90403
(2002).

\bibitem{Theorydiff} J. Javanainen and S. M. Yoo, Phys. Rev. Lett. {\bf
76}, 161 (1996); Y. Castin and J. Dalibard, Phys. Rev. A {\bf 55}, 4330
(1997).

\bibitem{Klausdiff} K. M\o lmer, Phys. Rev. A {\bf 55}, 3195 (1997);
J. Mod. Opt. {\bf 44}, 1937 (1997).


\bibitem{Levitt} M. H. Levitt, Prog. NMR Spectrosc. \textbf{18}, 61 (1986).

\bibitem{latest} Physics News Update, Number 663 {\bf 1}, November 25, 2003;
http://www.aip.org/enews/physnews/2003/split/663-1.html.

\end{thebibliography}
\end{document}